\documentclass[pre,twocolumn,showpacs]{revtex4}

\usepackage{graphicx}

\begin{document}

\title{Dynamics of polymer chain collapse into compact states}

\author{D. C. Rapaport}
\email{rapaport@mail.biu.ac.il}
\affiliation{Physics Department, Bar-Ilan University, Ramat-Gan 52900, Israel}

%\date{\today}
\date{July 14, 2003}

\begin{abstract}

Molecular dynamics simulation methods are used to study the folding of polymer
chains into packed cubic states. The polymer model, based on a chain of linked
sites moving in the continuum, includes both excluded volume and torsional
interactions. Different native-state packing arrangements and chain lengths are
explored; the organization of the native state is found to affect both the
ability of the chain to fold successfully and the nature of the folding pathway
as the system is gradually cooled. An order parameter based on contact counts
is used to provide information about the folding process, with contacts
additionally classified according to criteria such as core and surface sites or
local and distant site pairs. Fully detailed contact maps and their evolution
are also examined. 

\end{abstract}


\maketitle

\section{Introduction}

Several decades of protein folding simulation have produced a substantial body
of knowledge about processes governing molecular collapse from a random
denatured state to a well-ordered and uniquely defined ground state
\cite{bro88,dil95,sha97,bro98,pan00}. Simulations addressing detailed
representations of relatively large molecules over extended time periods lie
close to, or beyond, the capabilities of currently available computers;
consequently, a variety of simplified models designed to capture particular
aspects of the molecular behavior have been formulated. These models eliminate
much, or even most of the molecular detail, thereby achieving a major reduction
in the amount of computation required for evaluating the interactions that
govern the behavior. Further reduction in effort is gained by replacing
Newtonian dynamics by one of several forms of Monte Carlo sampling procedure,
and the configuration space available to the molecule is generally reduced
substantially by discretizing the problem and confining the molecule to the
sites of a regular lattice \cite{soc94,dil95,din96}. After imposition of these
and other approximations the connection between the models and the original
problem is at best tenuous, so that it is not always obvious if and how
conclusion arising from such models relate to real proteins.

Although a full dynamical simulation of a relatively detailed representation of
even a small protein (or part of a protein) remains a major computational
undertaking \cite{dua98}, there is no reason why simplified models cannot be
studied in the continuum with molecular dynamics (MD) methods. The purpose of
the present paper is to extend an earlier study of this kind \cite{rap02}, that
dealt with helix formation, to address the formation of a series of more
complex native structures; the expectation is that through such simulations
information will emerge about how structural variation influences the folding
process.

The native structure chosen for study is the cube. The molecule itself is
represented as a linear chain, free to move in the continuum (subject to
specified internal degrees of freedom), and the interactions are chosen in such
a way that, in the native low-energy state, the chain is folded so that its
sites form a perfectly ordered cubic array. There are numerous ways that a
linear path can wend its way through all the sites of a cubic lattice of a
given size; from this collection, four distinct but conveniently described
paths have been selected for study. Cubic conformations are not known to occur
in real proteins, but this is not an issue since the purpose of the study is to
examine the effect of the variation of ground-state configuration on folding
behavior. As will become apparent subsequently, the difference between these
states directly impacts both the likelihood of entanglement (misfolding) and
the rate of convergence to the ordered state (correct folding). This difference
is at least partly attributable to the dissimilar contact patterns: adjacent
sites in the native state involve different proportions of nearby and more
distant (or local and nonlocal) sites when measured in terms of backbone
separation.

In dealing with highly detailed protein models, there is a problem establishing
that the free-energy minimum of the overall potential function really
corresponds to the known native conformation; given the present state of the
art this is practically impossible, owing to the way potentials are developed
and the limited spatial resolution of experimental structure determination. The
alternative approach, based on a highly simplified description, preserves the
underlying tenet of protein structure theory, namely that the primary sequence
determines structure, but is sufficiently economical to allow complete folding
pathways to be studied. Moreover, the computational efficiency allows the
dynamics of not just a single folding process to be followed, but an entire
ensemble of systems can be examined, leading to a representative sample of the
kinds of behavior that can arise; the importance of extensive sampling cannot
be overstated, since with a very small number of samples it is impossible to
determine what is in fact ``typical'' behavior.

One reason for choosing a cubic form for the native state is that a
considerable body of work exists on chains that fold into this overall shape
(see \cite{din96} and references therein). The difference is that all such work
involves lattices, primarily using various Monte Carlo techniques; calculations
of this type are carried out at a constant nominal temperature (the temperature
is a parameter of the Monte Carlo procedure used in deciding whether to accept
randomly generated trial configurations, and it determines the ability to cross
potential-energy barriers). In lattice studies, the potential energy is a sum
over contributions from chain sites that are in contact because they occupy
adjacent lattice sites; the approach to choosing the potential and deciding
which site pairs should be encouraged to form contacts depends on the model,
and typically the interactions are chosen in such a way that the native state
has a substantially lower potential energy than any other state, even those
corresponding to different cube packings (in some of these studies,
interactions are selected so that even in the ground state a fraction of
adjacent site pairs would prefer not to be neighbors, thereby allowing a
certain amount of frustration to be incorporated into the design). The folding
behavior has been found to depend on the gap between this global energy minimum
and the other local minima corresponding to compact but misfolded
configurations. Much of the lattice work has dealt with cubes having three
sites per edge, corresponding to relatively short chains of length 27. When
chains of length 125 were studied by similar means \cite{din96} it was noted
that results based on shorter chains were subject to finite-size effects; in
particular, the energy gap was a necessary but not sufficient condition for
folding, and the presence of a set of core sites, nonexistent in 27-mers,
played a key role in the folding.

While it is possible to debate the merits of Monte Carlo relative to MD, there
is at least one difference that is particularly important for folding studies.
In lattice-based Monte Carlo, the spatial discretization imposes restrictions
on the permissible internal rearrangements: configuration changes typically
involve the displacement of a single chain site, or a very small number of
adjacent sites, as in crankshaft-type motion, and all displacements are based
on jumps between lattice sites. There is no provision for collective movement
of more extended subunits, a mode of reorganization likely to dominate once a
chain has reached an even moderately compact state, and clearly such
restrictions will have an impact on the chain ``dynamics''. The MD approach is
free from any limitations of this kind and, at least in principle, aims at a
realistic representation of the dynamics.

The goal of the present paper is to extend the approach used previously
\cite{rap02} for modeling helix collapse to the study of cubic configurations,
with emphasis on how the folding process is affected by the way the chain is
arranged inside the cube. The main focus of the earlier paper was on the
ability of a chain having a known native state to actually fold into that
structure, within the course of a simulation of reasonable duration; the
obvious extension of the work is to consider other configurations with
different structural features. A helix involves contacts between sites
relatively closely spaced along the chain, whereas cubic structures involve
varying mixtures of local and nonlocal neighbors, a distinction that is likely
to affect the folding process. The following sections outline the techniques,
insofar as they differ from the earlier work, and then proceed to a discussion
of the results and their implications.

\section{Methodology}

The model follows the approach described previously \cite{rap02} in which the
chain is represented as a series of sites (or masses) linked by bonds of
constant length, and the angles between successive bonds are assigned fixed
values consistent with the desired native ground state; the only internal
degrees of freedom are the dihedral angles that describe the relative
orientations of next-neighbor bond pairs projected in a plane perpendicular to
the bond between them. The interactions included in the model involve a
soft-sphere repulsion between sites (the short-range, repulsive part of the
Lennard-Jones potential) that is responsible for the excluded volume of the
chain, and a torsional potential acting along each bond (except for the first
and last bonds) whose minimum corresponds to the dihedral angle value in the
native state. Aside from the different bond and dihedral angles as described
here, all sites are identical. While both of these interactions appear in more
realistic protein potentials, the present model does not include any of the
site-specific pair interactions that are principal component of potentials used
in more detailed protein studies; since there are no direct interactions
between sites, any contact preferences exhibited by individual sites are
entirely due to the torsional interactions that act along the backbone.
Additionally, since the presence of a solvent would slow the computations
substantially, the chain resides in a vacuum (a characteristic of some more
detailed simulations as well); this also avoids the need to decide on the level
of detail used to describe solvent effects. At a qualitative level, omissions
such as these should not prevent simplified models from providing insight into
the underlying dynamics of folding.

The MD approach used for the simulation involves recursive techniques for
dealing with the internal degrees of freedom, while the integration of the
equations of motion (for translational, rotational and internal motion) is
based on the leapfrog algorithm; the technical aspects of computations of this
kind have been addressed at length elsewhere \cite{rap02,rap03} so that the
details need not be repeated here. The chain is initially heated to
comparatively high temperature to produce a random configuration, and is then
cooled very slowly by reducing the kinetic energy by a constant factor at
regular time intervals. The first stage of the cooling process is designed to
extract energy from the chain and leave it trapped in a free-energy well --
preferably, though not always, one from which the native state is accessible --
from which escape is unlikely; the second stage is intended to freeze the
configuration by eliminating, or at least weakening, the soft modes that
correspond to the low-frequency collective motion present even at low
temperature, so that measurements of the properties of the configuration
eventually reached are relatively free from the effects of thermal fluctuation.

As in the helix studies \cite{rap02}, the cooling rate is chosen empirically,
and is dependent on chain length. In order to allow comparison of the different
native-state chain arrangements, the same cooling rate is used in all cases.
The rate value is chosen to differentiate clearly between the varying levels of
ability to find the correct folding pathway to the native states; cooling too
fast will allow none of the chains to fold properly, while excessively slow
cooling will result in a situation where there are practically no
unsuccessfully folded chains.

As an alternative (not explored here) to gradual cooling, the simulation could
be carried out at constant temperature, using appropriate MD techniques, with a
temperature value chosen to achieve a reasonably rapid folding rate while
avoiding the excessive entanglement responsible for misfolding. Such results
would produce a long-tailed (or open-ended) distribution of folding times, so
that ensuring adequate configuration sampling is likely to demand much longer
computation times than the approach based on cooling. The information produced
concerning the comparative folding abilities of different native configurations
and the behavior observed along the folding pathways should be similar,
although timescales will differ due to the way temperature is involved.

The initial angular velocities associated with the dihedral angles are randomly
assigned; changes to the seed used for initializing the random number generator
lead to entirely different folding scenarios. Since it is neither possible, nor
particularly meaningful, to describe at length the detailed histories of each
of the large number of runs carried out, time-dependent ensemble averages over
the collection of histories are evaluated. In some of the analyses described
below the results are divided into two groups, runs that produce successful
folding and runs that become trapped in a misfolded state; other kinds of
analysis apply to either the entire set of runs or just the successful folders.
While the results could be divided according to other criteria, that based on
successfully completed folding is the most directly relevant.

Chains of length 64 and 125 have been studied; these chains can be packed into
cubic states with four and five sites per edge, respectively. In terms of the
reduced unit of length, chosen here (as usual) to be the parameter $\sigma$ of
the Lennard-Jones interaction (which, for the soft-sphere case, is slightly
less than the sphere's effective diameter) the length of the fixed links
between neighboring chain sites is 1.3; this bond length is sufficiently short
to prevent self-intersection, while allowing physically larger molecules than
if adjacent spheres were actually touching. A chain of $N$ sites has $N - 1$
links and, since torsional motion is not associated with the two end links, $N
- 3$ internal degrees of freedom. Bond angles are fixed at either $\pi / 2$
(right angle) or $0$ (parallel), depending on the location in the native-state
cube; to ensure that all bonds are treated equivalently, torsional motion is
associated with bonds even when successive bonds are in the same direction. The
torsional potential is the same sinusoidal function used in the helix studies,
with the energy minimum located at the native-state dihedral angle for each
individual bond.

The runs are of length $10^6$ and $1.5 \times 10^6$ time steps for the $N = 64$
and $N = 125$ chains respectively; the size of the time step is 0.004 in
reduced units. Cooling is accomplished by scaling the instantaneous kinetic
energy every 4000 steps by factors of 0.97 or 0.98 for the shorter and longer
chains, a larger factor implying slower cooling. This results in a kinetic
energy fall from an initial value of approximately 4 (per degree of freedom, in
reduced units defined in terms of the soft-sphere potential) to a final value
of $2 \times 10^{-3}$ (the sum of potential and kinetic energy first becomes
negative about a tenth of the way through the run); by comparison, the
potential energy per dihedral angle is $-5$ in the ground state. A total of 400
runs are carried out for each type of native state and chain length;
measurements and configurational snapshots are recorded at suitable time
intervals.

\begin{figure}
\includegraphics[scale=1.]{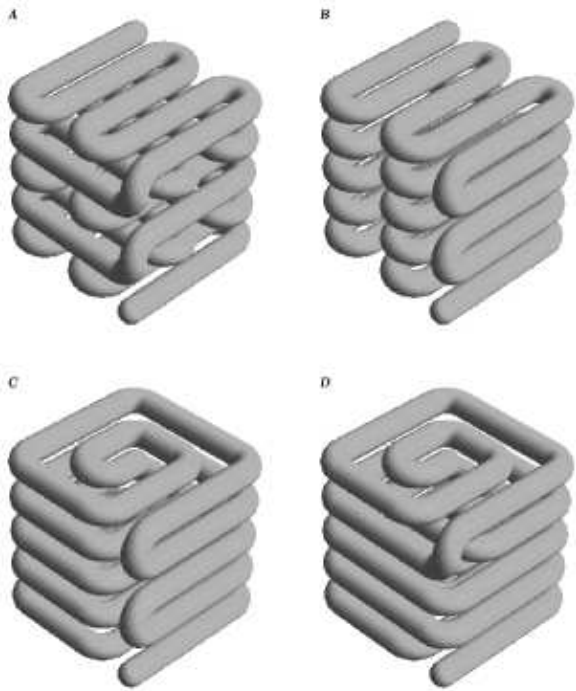}
\caption{\label{fig:pcube} Native-state configurations for cubes with five
sites per edge; for ease of visualization the chains are shown as continuous
tubes.}
\end{figure}

\begin{figure}
\includegraphics[scale=0.5]{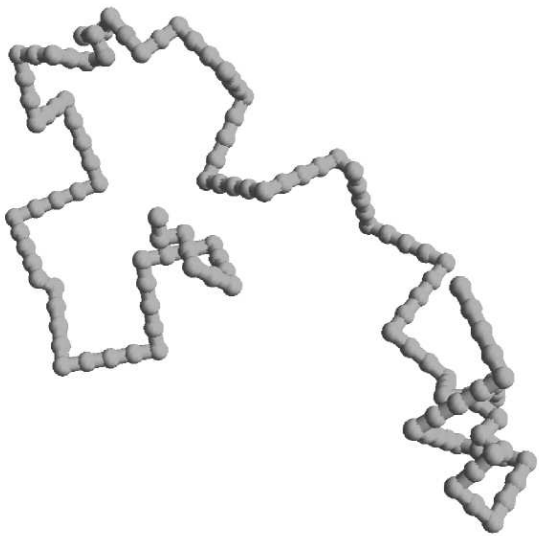}
\caption{\label{fig:prand} Typical random early state (type {\sl D}
configuration); the chain is shown as a sequence of linked spheres.}
\end{figure}

The four kinds of chain organization used for the native states are shown in
Fig.~\ref{fig:pcube}; here, for clarity, the chains appear as continuous tubes,
whereas in actual fact each chain consists of a series of closely-spaced
spheres joined by bonds of fixed length (an example of this representation is
shown subsequently). In the orientation shown, the cubes are filled one
complete horizontal plane at a time; the configurations shown are for $N =
125$, the organization for $N = 64$ chains is similar, after allowing for the
smaller cube size. The differences between the configurations relate to the
manner of filling the planes and the relationship between adjacent planes. Two
of the cases, labeled {\sl A} and {\sl B}, are based on zigzag (or
antiparallel) patterns, the other two, {\sl C} and {\sl D}, are spirals. The
other difference is the relation between the fill patterns of successive
planes: the two zigzag configurations consist of rotations ({\sl A}) and
reversals of the fill pattern ({\sl B}); the two spiral patterns consist of
spirals that rotate in alternate directions ({\sl C}) and spirals that continue
to rotate in the same direction ({\sl D}). In all cases, the chain terminal
sites lie on the outer faces of the cube. There are numerous other possible
patterns (some more readily characterized than others), in particular, patterns
that are not restricted to filling a single plane at a time (such as the
three-dimensional Hilbert curve), but the present assortment is already
sufficient for displaying a wide range of behavior. The filling sequences
create a distinction between ``secondary'' (each plane) and ``tertiary'' (the
entire cube) structural features which some alternative pathways through the
cube might not exhibit. Fig.~\ref{fig:prand} shows an early, essentially random
state from one of the runs after the chain has been heated to erase memory of
the initial state; here a ball-and-stick representation is used to show the
actual chain design.

\section{Results}

The analysis focuses on a selection of configurational averages and
distributions that characterize the time-dependent behavior of the chains as
folding progresses. Some quantities will turn out to be sensitive to the
organization of the native state, namely the way the cube is packed, others
less so. Although averaging merges the details of individual folding histories,
it does facilitate the extraction of key aspects of the behavior that are
contained in what is sometimes widely scattered data; the spread of the
distributions, where shown, provides an indication of the inherent variability
of the behavior.

\subsection{Folding success rate}

Ultimately, the only truly reliable test of whether a configuration has folded
successfully is based on visual examination, but while this technique has been
employed, it does not offer a reasonable approach for describing large sets of
configurations. Furthermore, owing to residual thermal fluctuations, the
separation between sites that are deemed to be within contact range will
usually exceed the value applicable to a perfectly formed cube, and this
flexibility must be accommodated by the analysis. A corresponding difficulty
does not arise in lattice studies, where the existence of a contact requires
the sites to be neighbors on the lattice in which the chain is embedded.

The automated decision scheme introduced here to classify the chain states
employs two defining parameters. The first is a factor $\phi$ multiplying the
bond length that establishes the maximum contact range. Visual inspection
suggests $\phi = 1.3$; this excludes essentially all false positives but can
underestimate the success rate by omitting some slightly misaligned but
otherwise correct configurations -- small thermal vibrations (generating
concertina-like modes that produce slightly nonparallel layers, or minor
deviations from planarity within the layers) are permitted energetically since
the cost is comparatively low. Raising the value to $\phi = 1.4$ improves the
situation by capturing most of these configurations, but allows a small number
of false positives. The second parameter is the minimal fraction of site pairs
that must satisfy the distance criterion for the chain to be regarded as
successfully folded. Here a value of 0.8 is used, and this fraction must appear
in at least one configuration during the final 120 time units (amounting to a
few percent) of the run; in practice, once this value is encountered, a high
contact fraction is normally maintained throughout the remainder of the run.

The fractions of chains of each type that fold successfully over the entire set
of runs are summarized in Table~\ref{tab:runs}. The results are ranked in
descending order; this order has also been used to assign labels to the
native-state configurations. The success rates for both values of $\phi$ are
shown; the difference is either zero or quite small. Thus, for practical
purposes, the results are insensitive to the choice of $\phi$ in this range,
and a value of $\phi = 1.3$ will be used when required in subsequent analysis.

For each configuration type the success rate is lower for the longer chains,
but for both cube sizes there is a significant, systematic variation that
depends on the organization of the native state. The most successful folder is
type {\sl A}, which consists (see Fig.~\ref{fig:pcube}) of successive zigzag
layers in a crossed alignment. Relatively similar success rates are achieved by
types {\sl B} and {\sl C}, which are the antiparallel (or reversed-direction)
zigzag layers and the reversed-direction spirals; the common feature of both
these configurations is that successive layers are mirror images of one another
(leading to prominent features in the contact maps discussed later). Finally,
the least successful folders are those of type {\sl D}, in which the spirals,
both increasing and decreasing, have all their turns in the same direction. The
fraction of successful folders as a whole depends on the cooling rate and, as
indicated earlier, the rates were chosen to allow differentiation between the
different chain types; the effect of alternative cooling rates is demonstrated
in \cite{rap02} for the case of helix folding, and cubes respond in a similar
manner.

\begin{table}
\caption{\label{tab:runs}Folding success rate for the different native-state
configurations and chain lengths ($N$); the results are based on a
contact-range factor $\phi = 1.3$ (the results for $\phi = 1.4$ are shown in
parentheses).}
\begin{ruledtabular}
\begin{tabular}{ccccc}
& Configuration & $N = 64$ & $N = 125$ & \\
\hline
& {\sl A} & 0.92 (0.92) & 0.81 (0.86) & \\
& {\sl B} & 0.78 (0.81) & 0.48 (0.53) & \\
& {\sl C} & 0.80 (0.80) & 0.40 (0.40) & \\
& {\sl D} & 0.51 (0.53) & 0.11 (0.12) & \\
\end{tabular}
\end{ruledtabular}
\end{table}

\begin{figure}
\includegraphics[scale=1.2]{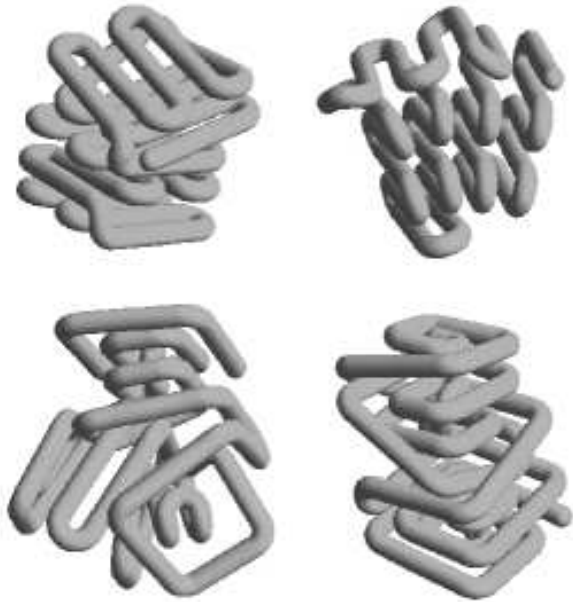}
\caption{\label{fig:pbad} Examples of misfolded states, one for each type of
configuration.}
\end{figure}

\begin{figure}
\includegraphics[scale=0.7]{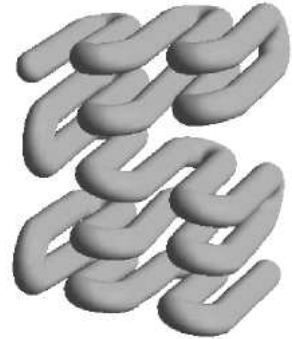}
\caption{\label{fig:pflex} A successfully folded configuration (type {\sl B})
subject to the effects of thermal vibration.}
\end{figure}

Fig.~\ref{fig:pbad} shows examples of misfolded states, one for each chain
type; much of the correct native structure is present, but a very small number
(as small as one) of incorrect twists leads to the wrong overall organization.
Visual examination of numerous final states leads to the conclusion that while
all correctly folded chains are alike, each incorrectly folded chain is
incorrect after its own fashion. One could attempt to classify incorrect folds:
in the helix study, most misfolds were due to a single incorrect twist, while
here, an example of a clearly identifiable misfold involving just a single
twist is the case in which an outer layer is rotated out of alignment and
jammed against one of the other faces of the cube. However, it is not apparent
that any useful purpose would be served by a systematic search among the
misfolded states for common structural motifs. Residual thermal vibration was
the reason given earlier for failing to identify states as correctly folded; an
example of a folded configuration representing a borderline case is shown
Fig.~\ref{fig:pflex}.

\subsection{Energy and radius of gyration}

The most familiar global properties of flexible chain molecules are the
internal energy and the radius of gyration (another quantity familiar from
chain-configuration studies, the mean-square end-to-end separation, does not
convey sufficient information to be useful for analyzing folding). The distance
of each of these quantities from the known native-state value provides an
averaged measure of the deviation of the current chain state from the ordered
configuration, without any attempt at characterizing the precise form of this
deviation. These quantities, unlike some more specific measures described
later, also correspond to properties that, at least in principle, are
measurable (by calorimetry and light scattering) in the laboratory.

\begin{figure}
\includegraphics[scale=0.8]{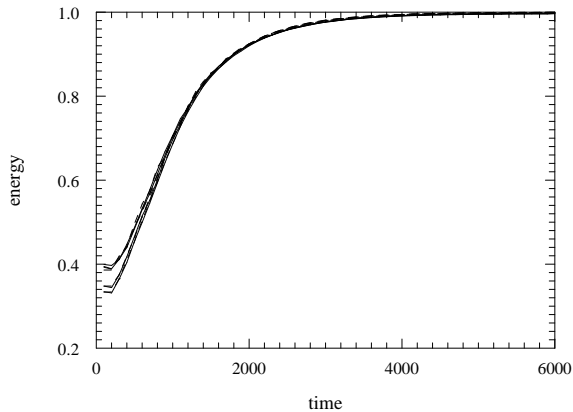}
\caption{\label{fig:peng} Mean internal energy (normalized) {\sl vs} time for
successful and unsuccessful folders of all configuration types for chains of
length 125 (reduced units are used).}
\end{figure}

The graphs of normalized internal energy (the only interactions incorporated in
the model, apart from excluded volume which makes a very small contribution,
are the torsional terms) are shown in Fig.~\ref{fig:peng} for $N = 125$; since
the energies are negative the magnitudes are shown. The overall average for
each of the four sets of configurations is plotted as a function of time, with
separate curves used to distinguish the averages for successful and
unsuccessful folders (using the definition of success given above). It is clear
from the graph that there is very little to separate the different sets of
data, implying that energy is not a useful distinguishing factor for these
studies. Examination of the full energy distributions (not shown) reveals the
spread in values to be very narrow. As will become apparent subsequently, the
internal energy approaches the limiting value of the folded state well before
other quantities achieve convergence.

\begin{figure}
\includegraphics[scale=0.8]{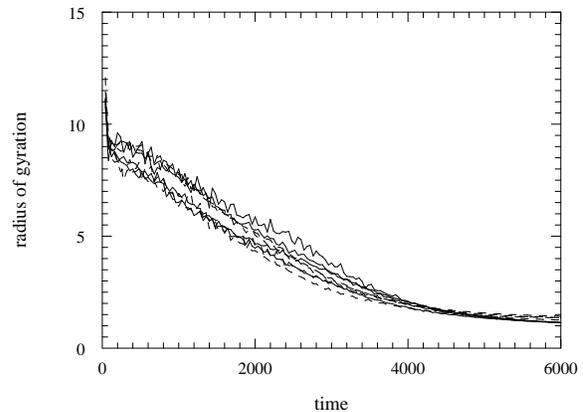}
\caption{\label{fig:prgyr} Mean radius of gyration (normalized) {\sl vs} time
for successful (solid curves) and unsuccessful (dashed) folders of all
configuration types for chains of length 125.}
\end{figure}

An overall measure of the compactness of the configuration is provided by the
radius of gyration; it is defined as the root-mean-square distance of sites
(assuming all to have the same mass) from their mutual center of mass (if the
shape deviates significantly from spherical -- or cubical -- symmetry, then it
might be necessary to consider the individual moments of the mass distribution,
but this is not the case here). The mean radius of gyration is shown in
Fig.~\ref{fig:prgyr}, again with separate curves for successful and
unsuccessful folders of each configuration type. The results are normalized
relative to the correctly folded collapsed state, which for a bond length of
1.3 has the value 10.14 for a packed cube of size five (6.34 for size four).
The results exhibit some degree of spread over much of the run duration,
reflecting different collapse rates, but the final results are very closely
spaced.

A common feature of the results in Fig.~\ref{fig:prgyr}, for all configuration
types but more pronounced in some cases, is that the unsuccessful folders
appear to collapse somewhat more rapidly during the initial stage of folding,
only to be overtaken later on by the chains that do manage to fold correctly.
There is very little difference in the final mean values for the different
configurations and folding outcomes. The actual distributions of values (not
shown) are fairly narrow, but do reveal additional details; for example, the
distribution at the end of the runs for type {\sl C} configurations consists of
a pair of narrow partially overlapping peaks, although the combined peak width
is no greater than for type {\sl D} with only a single peak. Since it follows
from Fig.~\ref{fig:prgyr} that the range of actual (unnormalized) mean values
is less than the site diameter it is apparent that, as with the internal
energy, the radius of gyration is of little help in differentiating between
sets of correctly and incorrectly folded states (the more extreme cases of
misfolding make only a small contribution to the mean). Though the radius of
gyration converges less rapidly than the energy, it is already close to its
limiting value at a time when other chain properties (see below) are still
undergoing substantial change. This initial collapse to a relatively compact
state, followed by a (sometimes successful) search for the route to the native
folded state, is also a feature of lattice Monte Carlo studies \cite{din96};
however, given the continuity of configuration space in the MD approach, it is
likely that MD will be more effective in rearranging a chain once it has
reached a semi-compact state.

\subsection{Contact formation}

\begin{figure}
\includegraphics[scale=1.26]{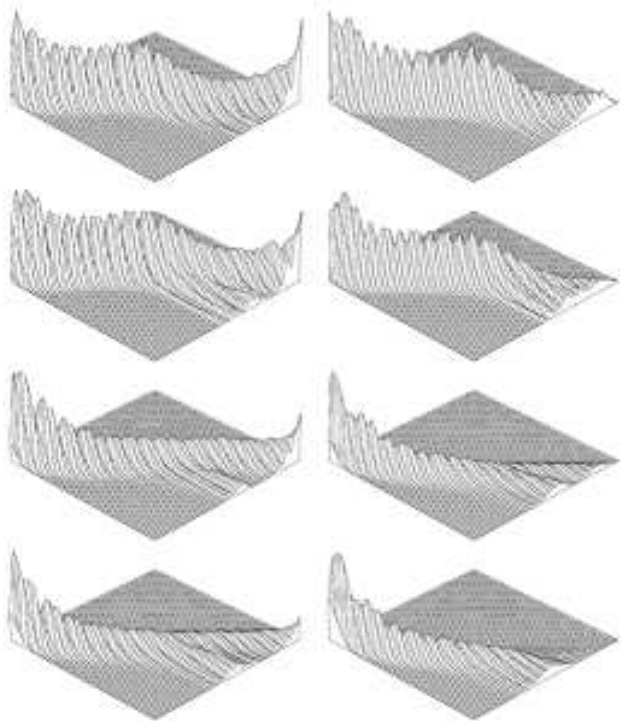}
\caption{\label{fig:pbfrac} Contact pair fraction $S$ {\sl vs} time for all
configurations ({\sl A} to {\sl D} ordered vertically) and cube sizes (4 on the
left, 5 on the right); in each of these and subsequent plots of a similar kind,
the time axis runs along the lower left edge of the grid, the value of the
quantity whose distribution is being measured along the right edge, and the
distribution at each instant is normalized.}
\end{figure}

The study of properties based on the local environments of the chain sites
provides an alternative to the global measures of energy and conformation
considered above. An order parameter $S$ that measures the proximity of a
configuration to its native state can be defined in terms of the fraction of
site pairs forming contacts in the native state that lie within contact range
-- with the separation factor $\phi$ introduced earlier used to determine which
pairs qualify (for a chain whose native state is a cube with $n$ sites per
edge, the maximum number of contacts is the number of lattice bonds $n (n - 1)
(3 n - 2)$ minus the number of chain bonds $n^3 - 1$). The time dependence of
$S$ is shown in Fig.~\ref{fig:pbfrac}. Here, and subsequently, surface plots
are used to display these time-varying distributions in compact fashion. The
fraction of successful folders corresponds to the area under the peak in the
region close to unity near the right corner of the grid; unsuccessful folders
contribute to lower values. Clearly, the behavior of $S$ depends not only on
chain length but also on the cube packing. In some cases, the $S$ distribution
becomes bimodal in the course of the folding process, with the two peaks
corresponding to chains that do and do not manage to fold successfully ($S
\approx 0.8$ is the minimal value exhibited by a folded chain, in which a
fraction of the pairs that should be classified as being in contact is
temporarily out of range due to thermal fluctuations); in other instances, a
single broad peak encompasses both folded and misfolded configurations, and its
position depends on the configuration type.

\begin{figure}
\includegraphics[scale=1.]{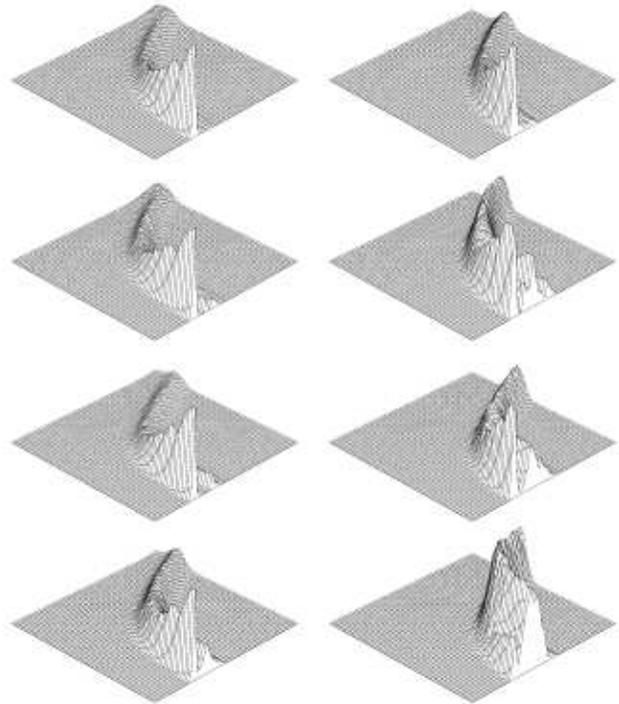}
\caption{\label{fig:pbsep} Separation of site pairs that are in contact when in
the native state {\sl vs} time (the sequence of plots is the same as in
Fig.~\ref{fig:pbfrac}).}
\end{figure}

An alternative approach to describing the degree of contact formation that
avoids the need for the factor $\phi$ is based on measuring the separation
distribution of site pairs that should lie within contact range. The results
are shown in Fig.~\ref{fig:pbsep}; in those cases where multiple peaks occur,
it is the narrow peak located at the lowest value that corresponds to correctly
positioned neighbors, with peaks at larger separations showing the distance
distribution of those site pairs that are unable to position themselves
correctly; the graphs include all chains, both folders and non-folders, with
both kinds contributing to the various peaks, although in different
proportions.

\begin{figure}
\includegraphics[scale=1.26]{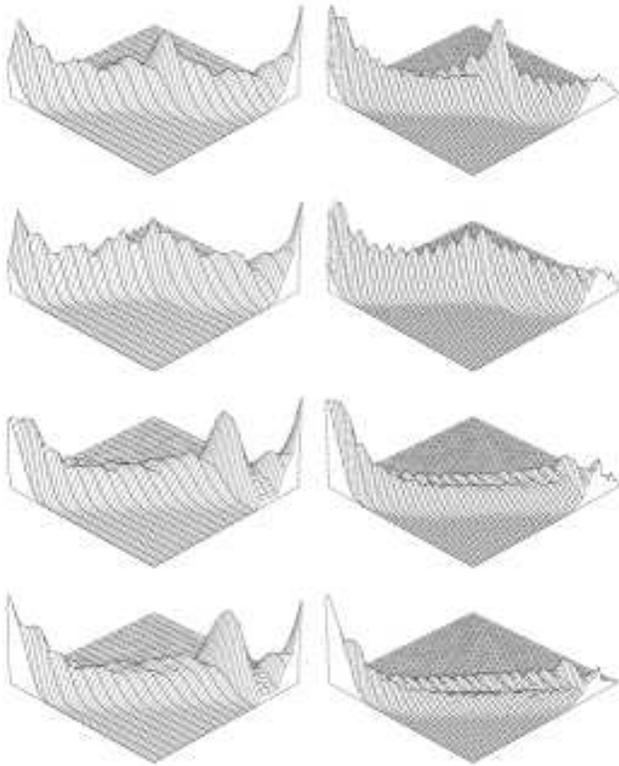}
\caption{\label{fig:pbcor} Contact fractions for core and surface sites {\sl vs}
time for chains that fold successfully (plots are organized as before).}
\end{figure}

Contact formation between particular site pairs can be selectively studied by
classifying the sites into different categories and evaluating the average
behavior for each category. One such classification, used previously in lattice
studies \cite{din96}, distinguishes sites belonging to the exterior surface
faces of the cube from those of the interior core. Although the design of the
native states of the chains considered here does not assign any preferred role
to such a core, it does not preclude analysis of the results from such a
perspective. Chains that fill cubes of size four have a mere eight sites in the
core (12.5\% of the sites), but chains filling cubes of size five have a more
substantial core of 27 sites (21.6\%). Fig.~\ref{fig:pbcor} shows the contact
fractions for chains that are able to fold successfully (the two sets of
contact distributions -- for core and surface sites -- are normalized
separately and then combined in each plot); with the longer chains, it is
apparent that for configurations {\sl C} and {\sl D} contacts form, on average,
at different rates and the core sites tend to make contact with their correct
neighbors earlier (separate plots of the distributions establishes this order).
Such behavior is indicative of structure (in an averaged sense) nucleating in
the interior and propagating outwards, but it is not apparent for all chain
types and lengths.

\begin{figure}
\includegraphics[scale=1.26]{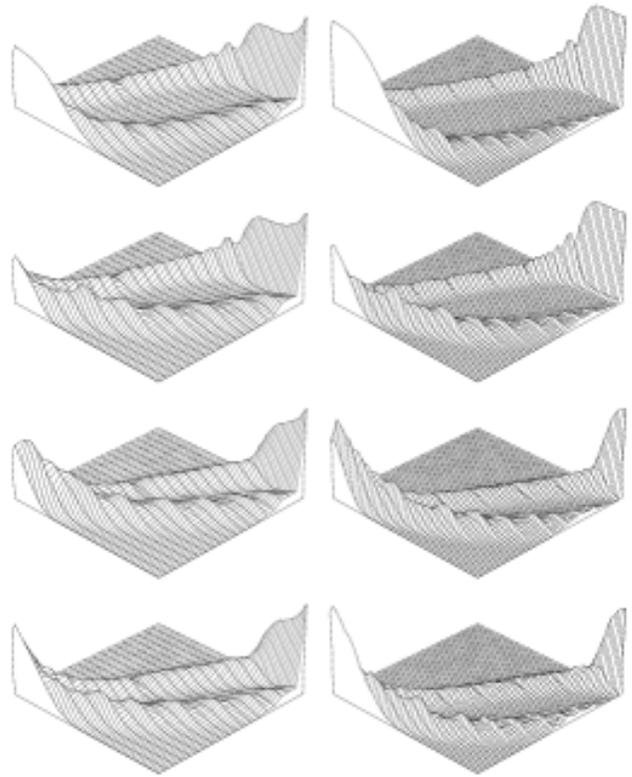}
\caption{\label{fig:pblay} Contact fractions for site pairs within and between
layers {\sl vs} time for chains that fold successfully.}
\end{figure}

An alternative to making the distinction between core and surface sites is to
classify the individual contacts according to their role in the structure. All
the chain configurations considered in the present work share the common design
characteristic of a native structure built from a series of planar layers. This
suggests dividing the contacts into two classes, those between sites within the
same layer (each layer can be regarded as a secondary structure element) and
those between sites in adjacent layers (forming the tertiary structure). The
results, shown in Fig.~\ref{fig:pblay}, reveal that in all cases (the two
contact distributions have again been combined in each plot) the contacts
within layers tend to form first, followed, after a delay, by contacts between
layers. Furthermore, the formation of contacts within layers is essentially
complete, and the missing contacts are those that should have appeared between
the layers. The detailed distributions, and especially the elapsed time between
the formation of specific fractions of intra- and inter-layer contacts, vary
substantially with configuration type (and, of course, with chain length).

The tendency for intra-layer contacts to form first, and in some instances for
the core sites to form contacts earlier, are typical of the kinds of general
observations that can be made concerning the folding pathways of simplified
models. There is no component in the potential energy function that directly
prefers any particular feature over any other (unlike the helix pairs of
\cite{rap02} where the potential favored the prior formation of individual
helices and only subsequently their parallel alignment), so that this behavior
is something that emerges spontaneously from the model. Owing to the wide
variation in possible pathways, itself a consequence of the many degrees of
freedom of the chain and the resulting high dimensionality of configuration
space, it is not obvious how to further categorize the folding pathways at a
higher level of resolution (by, for example, introducing appropriately defined
``reaction coordinates''). In particular, it does not seem possible to
establish the existence of ``funnels'' in configuration space that are
traversed by a relatively large proportion of pathways; whether such a concept
is useful (except in certain very specific cases) is questionable, indeed there
could well be so many funnels that they span a substantial portion of
configuration space. 

\subsection{Contact maps}

The contact map provides a highly detailed but concise summary of the proximity
of a configuration to its native state by showing which site pairs are within
contact range; it also reveals patterns in the distribution of contacts among
chain sites as a function of their backbone separation and among groups of
sites at particular backbone locations. It is also possible to forgo these
details and study how the contact formation rate depends on backbone separation
alone, without taking into account which sites are involved; such results
reveal, not unexpectedly, that nearby (more ``local'') sites generally form
contacts earlier and have a higher overall success rate (there is the
occasional exception). This classification, however, ignores features that have
already been found to be significant, such as contacts within and between
layers. Although the full contact map is not subject to this shortcoming, it
does have the opposite problem of an excess of detail, and because sample sizes
are smaller due to each pair of sites being treated separately, the results are
subject to increased statistical noise.

The general arrangement of a contact map is that both axes correspond to the
indices of the sites along the backbone; if a pair of sites $(i, j)$ are within
contact range then a spot appears at the corresponding coordinates. Since the
plot is symmetric only half the spots need be shown; the backbone pairs can be
included for reference even though they are always present. It is possible to
extend this scheme for representing the configuration to produce a reasonably
reliable, low-temporal-resolution history of the folding process; this is
accomplished by replacing the binary (pair/nonpair) values in the contact map
by continuous values representing the fraction of configurations sampled in
which pairing occurs (including only those sites that are actually paired in
the native state) over the course of folding, and using either color or
greyscale coding to denote the value. On the assumption that the contacts that
appear more frequently are those that develop earlier along the folding
pathway, the history is contained in this generalized version of the contact
map; a considerably less concise alternative would be a sequence of binary maps
corresponding to different times during the simulation.

\begin{figure}
\includegraphics[scale=0.88]{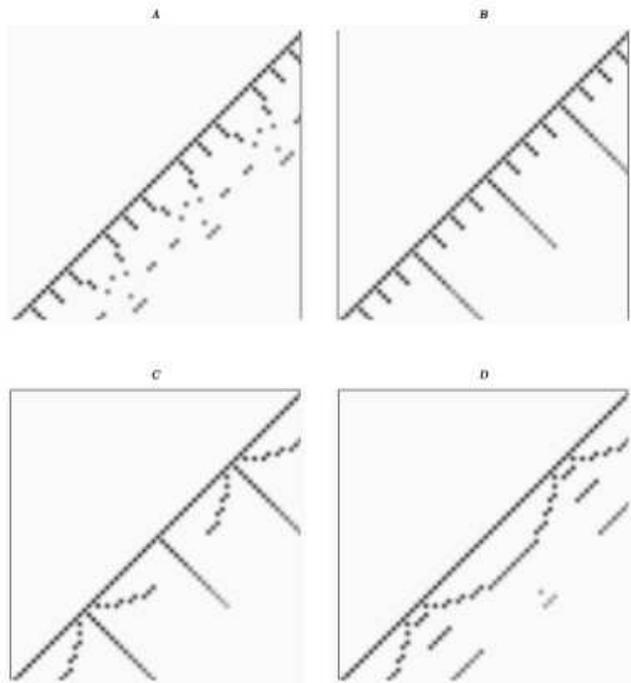}
\caption{\label{fig:pcont} Generalized contact maps for cubes of size four
(only chains that fold successfully contribute); shading is used to indicate
the folding history, with darker spots corresponding to contacts that form
earlier (the main diagonal represents the backbone neighbors).}
\end{figure}

The contact maps shown in Fig.~\ref{fig:pcont} incorporate two different kinds
of information. The first is the actual pattern of contacts appearing in the
ground state of the chain, which is indicated by the presence of spots
irrespective of their shading; only the results for the shorter chains are
shown because for longer chains the details are too fine to be reproduced in
these small figures. Certain structural features lead to recognizable patterns
in the contact map, and while particular motifs appear in more than one map,
the overall patterns associated with the four configurations are seen to be
very different (there is also a systematic dependence on chain length).
Contacts involving nearby (in terms of backbone separation) pairs appear close
to the diagonal; identifiable structural elements in the native state, such as
zigzag layers, spirals and reversed adjacent layers, correspond to specific
spot patterns.

The other category of information present in the generalized contact map is a
concise representation of the folding history of the chain ensemble. Since the
greyscale representation embodies the averaged folding history, the time at
which a contact typically appeared can be inferred from the darkness of the
corresponding spot (the use of color increases the temporal resolution, so that
color plots were also used in examining the behavior). What sort of information
can be extracted from maps of this kind?

Given that in the present model the contacts themselves do not contribute
directly to the interaction energy, but only reflect the organization dictated
by the torsional potential, it is not surprising that the maps reveal that
contacts between nearby sites along the backbone tend to appear earlier (there
are, for example, no attractive forces acting directly between distant sites
that could compete against this ordering). The behavior is not always
monotonic, so that in the longer, linear spot sequences perpendicular to the
main diagonal, a feature associated with configurations of type {\sl B} and
{\sl C}, both of which have reversed adjacent layers, there are instances of
more distant backbone sites pairing (on average) prior to those that are
nearby. One could attempt to correlate specific features in the contact maps
with the ability of the chains to fold successfully; while there are particular
details that are likely to be relevant, such as the presence or absence of
larger groups of spots at various distances from the main diagonal, the data
are probably insufficient to allow any firm conclusions. What is apparent here
is that the best folder, type {\sl A}, has the least amount of structure of
this kind, {\sl B} and {\sl C} are similar, while {\sl D}, the poorest folder,
has the most structure (these observations also apply to the longer chains).
The inclusion of additional native-state structures in the study might help
elucidate the relationship between contact patterns and foldability.

\section{Conclusions}

The focus of this paper has been on the folding ability of model chain
molecules with well-defined native states. While each of the native-state
configurations is cube filling, the paths taken by the chains through the cubes
are different for each configuration, and this difference -- which in some
sense determines the ``accessibility'' of the native state -- has a strong
influence on the outcome of the folding process. Unlike the requirement of many
lattice studies, the native-state energy is not markedly lower than for compact
misfolded states; nevertheless, some of the configurations already achieve high
folding success rates under the conditions of the simulation, and the results
for the less successful cases could be further improved by slower cooling. The
key conclusion is, therefore, that if a unique native configuration exists that
is consistent with the interaction potential, the chain is able -- both in
principle and in practice -- to find it. The difficult problem, in the context
of more realistic models, is how to formulate such a potential. 

The relevance of the study of cubic conformations, despite the fact that such
systems have no counterpart in nature, is the existence of well-defined,
unambiguous ground states. Additionally, the cubic native state is able to
accommodate a combination of both short- and long-range structural features.
The fact that there have been extensive studies of cubically packed chains
using other simulational approaches is yet another motivating factor. The
advantage of the continuum MD approach is that efficient collective
reorganization can occur, even in relatively compact states, as opposed to the
discrete sets of moves provided by the various Monte Carlo approaches; the
absence of configurational restrictions imposed by an underlying lattice allows
for continuous conformational change, an important capability for
partially-collapsed chains.

The most obvious limitations of the model, in its present form, is that the
only interactions included, aside from excluded volume, are of torsional type,
and that no solvent is included. Neither shortcoming is difficult to overcome
in principle. Additional interactions could be included if there was a specific
reason to do so, although consistency of the overall potential with the desired
native state would have to be assured. The inclusion of an inert solvent,
modeled using discrete particles, would slow the dynamics substantially and
require additional computation to handle the extra degrees of freedom; it would
also be possible to incorporate specific chain--solvent interactions (to mimic
hydrophobicity for example), but this, too, would be a complicating factor that
the deliberately simple design of the present model is aimed at avoiding.

Even with these simplification, the model has been shown to display a range of
physically interesting and relevant kinds of behavior. All the observations are
based on a chain model whose interactions have been designed without any
specific folding scenario in mind (aside from the chosen native state);
nevertheless, there are certain identifiable aspects of the mean behavior that
merit attention, such as the tendency in certain configurations for the core
sites to form contacts earlier, for layers to become organized first, and for
the correct folding pathway to be correlated with a slower overall collapse (as
measured by the radius of gyration). Such features correspond to generic
behavior, so that simple models of this kind can be used as reference systems
when investigating more highly refined designs. If additional features are able
to produce little more than already occurs in the simplest of models, then they
clearly fail to advance the state of the art.

What lessons emerge from these results insofar as protein folding is concerned?
Each of the cases studied has a single minimum (free-) energy collapsed state
and a variety of misfolded states with very similar energies. Thus there is no
global minimum state that is well-separated from the many local minima, and on
the basis of energy considerations alone it is not possible to reach any
conclusions about the folding capability of each kind of chain. The intuitive
but non-quantifiable notion of accessibility -- the ability of the chain to go
where it needs to be -- plays an apparent role, since it reflects the
robustness of the chain in resisting the consequences of incorrect
entanglement; partially entangled states occur often along a folding pathway,
but their effect is transient if escape is possible. It is clear from the
present results that even the simple chain models studied here differ greatly
in this respect.

\begin{acknowledgments}
This work was partially supported by the Israel Science Foundation.
\end{acknowledgments}

\bibliography{foldcube}

\end{document}